\documentclass[dvips]{imsart}

\RequirePackage[OT1]{fontenc}
\usepackage[dvips]{graphicx,color}
\usepackage{amsthm,amsmath,amssymb,natbib}
\RequirePackage{hyperref}
\usepackage{natbib}
\usepackage{har2nat}

\arxiv{arXiv:0000.0000}

\startlocaldefs
\numberwithin{equation}{section}
\theoremstyle{plain}
\newtheorem{thm}{Theorem}[section]
\newtheorem{prop}{PROPOSITION}[section]
\newtheorem{lema}{LEMMA}[section]
\theoremstyle{definition}
\newtheorem{defin}{Definition}[section]
\newtheorem{note}{Remark}
\endlocaldefs

\newcommand{\chunk}[3]{#1_{#2:#3}}

\begin{document}

\begin{frontmatter}
\title{Robust estimation in time series with long and short memory properties}
\runtitle{Robust estimation in time series}

\begin{aug}
\author{\fnms{Vald\'erio A.}
\snm{Reisen}\thanksref{a,t2}\ead[label=e1]{valderionselmoreisen@gmail.com}}
\and
\author{\fnms{Fabio A.} \snm{Fajardo}\thanksref{b}\ead[label=e2]{fabio.molinares@ufes.br}}

\thankstext{t2}{The first author  gratefully acknowledge partial financial support from 
CNPq-Brazil.}

\runauthor{V. Reisen and F. Fajardo}

\affiliation[a]{DEST, PPGEA, University of Espirito Santo, Vit\'oria, ES, Brazil.}
\affiliation[b]{DEST, PPGECO, University of Espirito Santo, Vit\'oria, ES, Brazil.}

\address{DEST, UFES, Av. Fernando Ferrari, 514, Goiabeiras. Vit\'oria - ES - 
CEP 29075-910,\\
\printead{e1,e2}}


\end{aug}

\begin{abstract}
This paper reviews recent developments of robust estimation in linear time series 
models, with short and long memory correlation structures, in the presence of additive
outliers.  Based on the manuscripts \citeasnoun{fajardo:reisen:cribari:2009} and 
\citeasnoun{levy:boistard:moulines:taqqu:reisen:2011a}, the emphasis in this paper is 
given in the following directions; the influence of additive outliers in the estimation
of a time series, the asymptotic properties of a robust autocovariance function and a 
robust semiparametric estimation method of the fractional parameter $d$ in  
ARFIMA$(p,d,q)$ models. Some simulations are used  to support the use of the robust 
method when a time series has additive outliers. The invariance property of the 
estimators for the first difference in ARFIMA model with outliers is also discussed. In
general, the robust long-memory estimator leads to be outlier resistant and is invariant
to first differencing.
\end{abstract}




\end{frontmatter}

\section{Introduction}
Let $\{X_t\}_{t\in\mathbb{Z}}$ be a stationary time series with spectral density that
behaves like
\begin{equation*}
f_{X}(\omega)\sim h(\omega)\mid\omega\mid^{-2d},
\text{ as } \omega \rightarrow 0
\end{equation*}
where the spectral density $h(\omega)$ is a nonvanishing and continuously differentiable
function with bounded derivative for $-\pi\leq\omega\leq\pi$, and $d < 0.5$.

A well-known stationary parametric model with  the above spectral density is
the ARFIMA$(p,d,q)$ process, which is the solution of the equation
\begin{equation}\label{eq:arfima}
 X_{t}-\mu =(1-{B})^{-d}\eta_t, \text{ with } t\in \mathbb{Z},
\end{equation}
where $\eta_t = \frac{\Theta(B)}{\Phi (B)}\epsilon_{t}$ is the
ARMA$(p,q)$ process, $\mu $ is the mean (here it is assumed that
$\mu = 0$), $\Phi ({B})\equiv 1-\sum_{j=1}^{p}\phi_{j}{B}^{j}$,
$\Theta ({B})\equiv 1-\sum_{i=1}^{q}\theta_{i}{B}^{i}$ and $p$ and
$q$ are positive integers \cite{hosking:1981}. $\Phi ({z})$ and $\Theta ({z})$, with a
scalar $z$, are polynomials with all roots outside the unit circle
and share no common factors. $d$ is the parameter that holds the
memory of the process, that is, when $d \in (-0.5,0.5)$ the
ARFIMA$(p,d,q)$ process is said to be invertible and stationary.
Besides, for $d\neq 0$, its autocovariance decays at a hyperbolic
rate ($\gamma(j)= O(j^{-1+2d})$). For $d=0$, $d \in (-0.5,0)$ or
$d \in (0,0.5)$, the process is said to be short-memory,
intermediate-memory or long-memory, respectively. The long-memory
property is related to the behavior of the autocovariances, which
are not absolutely summable and the spectral density becomes
unbounded at zero frequency. In the intermediate-memory region,
the autocovariances are  absolutely summable and, consequently,
the spectral density is bounded.

The spectral density function of $\{X_t\}_{t\in \mathbb{Z}}$ is given by
\begin{equation*}
f_{X}(\omega)=f_{\eta}(\omega)\left[2\sin\,\left( \frac{\omega}{2} \right) \right]^{-2d},
\,\omega\in \lbrack -\pi ,\pi ]. 
\end{equation*}
$f_{X}(\omega)$ is continuous except for $\omega=0$ where it has a
pole when $d>0$. A recent review of the ARFIMA model and its
properties can be found in \citeasnoun{palma:2007} and 
\citeasnoun{doukhan:oppenheim:taqqu:2003}.

Many estimators for the fractional parameter $d$ in long-memory time series have already
been proposed in the literature. Among them are the semiparametric procedures, a group 
which includes a wide variety of estimators based on the Ordinary Least Square (OLS)
method. These procedures require the use of the spectral density parameterized within a
neighborhood of zero frequency. Some references on this subject include the works of 
\citeasnoun{geweke:porter-hudak:1983}, \citeasnoun{reisen:1994} and 
\citeasnoun{robinson:1995a}, among others. An overview of long-range dependence 
processes can be found in \citeasnoun{beran:1994} and 
\citeasnoun{doukhan:oppenheim:taqqu:2003}.

Time series with outliers or atypical observations  is quite common in any area of 
application. In the case where the data is time-dependent, several authors such as 
\citeasnoun{ledolter:1989}, \citeasnoun{chang:tiao:chen:1988} and 
\citeasnoun{chen:liu:1993b} have studied the effect of outliers in a time series that 
follows ARIMA models. In general, they have concluded that the parameter estimates of 
ARMA models become more biased when the data contains outliers. Similar conclusion is 
also observed when estimating the fractional parameter in ARFIMA models. The outliers 
cause a substantial bias in the differencing parameter 
\cite{fajardo:reisen:cribari:2009}.

An autocovariance robust function was proposed by \citeasnoun{ma:genton:2000}. The 
asymptotical properties of this function are studied by 
\citeasnoun{levy:boistard:moulines:taqqu:reisen:2011a}. The results presented in 
\citeasnoun{fajardo:reisen:cribari:2009}, 
\citeasnoun{levy:boistard:moulines:taqqu:reisen:2011a} and 
\citeasnoun{levy:boistard:moulines:taqqu:reisen:2011c} are the motivations of this 
paper. The impact of outliers in the estimation of ARFIMA models under different context
is here studied. The asymptotical properties of a  robust autocovariance function is 
discussed and some empirical examples are used to illustrate the usefulness of a robust
fractional parameter estimator. The invariance property of the estimator to the first 
difference is also empirically studied. The outline of this papers is as follows:
Section \ref{sec:efeito} discusses the model and the impact of the outliers in time 
series. Section 3 summarizes the main results related to the robust autocovariance 
estimator given in  \citeasnoun{levy:boistard:moulines:taqqu:reisen:2011a} and discusses
the robust estimation of the fractional parameter in the ARFIMA model. Section 4 
presents some empirical studies and an application is discussed in Section 5. Concluding
remarks and future directions are given in Section 6.

\section{The impact of outliers in time series}\label{sec:efeito}
Suppose  $x_1,\ldots,x_n$ is a partial realization of $\{X_t\}_{t\in \mathbb{Z}}$. 
Hence, the periodogram function is defined as
$I_{x}(\omega) = {(2\pi n)}^{-1} |\sum_{t=1}^{n} x_{t}e^{i\omega
t}|^{2}$. It follows that, when $d= 0$ in the ARFIMA model,
\begin{equation}
I_{x}(\omega) = 2\pi
f_{X}(\omega)\frac{I_{\epsilon}(\omega)}{\sigma^{2}_{\epsilon}} +
H(\omega)
\end{equation}
where $\mathbb{E}[|H(\omega)|^2] = O( \frac{1}{n^{2\xi}})$ ($\xi>
0 $) is uniformly in $\omega \in [-\pi, \pi]$ (Theorem 6.2.2 in
\citeasnoun{priestley:1981}) and $I_{\epsilon}(\cdot)$ is the periodogram of the 
residuals. From Equation \ref{eq:arfima} and Theorem 6.1.1 in
\citeasnoun{priestley:1981}, asymptotic sample properties of
$\frac{I_{x}(\omega)}{f_{X}(\omega)}$ are derived and they are
summarized as follows. If $\left\{\epsilon_{t}\right\}_{t\in \mathbb{Z}}$ are
normally distributed, for a fixed set of values of the Fourier
frequencies $\omega_j = \frac {2\pi j}{n}$, $j= 1,\ldots, \lfloor n/2\rfloor$,
where $\lfloor\cdot\rfloor$ means the integer part, asymptotically the set of variables
$\frac{I_{x}(\omega_j)}{f_{X}(\omega_j)}$ is independently distributed, each distributed
as $\frac{ \chi_2^2}{2}$. At $\omega = 0$ and $\pi$, the distributions are $\chi_1^2$ 
(for details see \citeasnoun{priestley:1981}). These asymptotic results for the 
periodogram lead to $\mathbb{E}\left[\frac{I_{x}(\omega_j)}{f_{X}(\omega_j)}\right]
\rightarrow 1$ and var$\left[ \frac{I_{x}(\omega_j)}{f_{X}(\omega_j)} \right]
\rightarrow (1 + \delta(\omega_j))$  as $n \rightarrow \infty$,
where
\begin{equation}\label{def:delta}
\delta(\omega_j)=1\textrm{ if } \omega_j = 0,\pi \textrm{ and } 0
\textrm{ otherwise}.
\end{equation}
The above results  establish the  unbiasedness and inconsistency
properties of $I_{x}(\omega_j)$.

Due to the singularity of $f_{X}(\omega)$ when $d > 0$, the
standard results of the asymptotic distribution of the periodogram
discussed previously can not be applied to $I_{x}(\omega_j)$ for
small and fixed $j$. \citeasnoun{hurvich:beltrao:1993}  showed  that $\lim
_{n \rightarrow \infty }
\mathbb{E}\left[\frac{I_{x}(\omega_j)}{f_{X}(\omega_j)}\right]$ depends on
$j$ and $d$, and exceeds unity for most $d\neq 0$ (\citeasnoun{kunsch:1986};
\citeasnoun{robinson:1995b}). For $ j\neq k$,
$\frac{I_{x}(\omega_j)}{f_{X}(\omega_j)}$ and
$\frac{I_{x}(\omega_k)}{f_{X}(\omega_k)}$ are correlated, and for a fixed
value $j$ and Gaussian processes, the limiting distribution of
$\frac{I_{x}(\omega_j)}{f_{X}(\omega_j)}$ is not exponential \cite{robinson:1995b}. 
That is, under the Gaussian assumption, \citeasnoun{hurvich:beltrao:1993} show that the
normalized periodogram $\frac{I(\omega)}{f_X(\omega)}$ is asymptotically distributed as
the quadratic form
\begin{equation}
 \frac{ \alpha_1}{2}\chi_{1} +  \frac{ \alpha_2}{2} \chi_{2}
\end{equation}
where $\chi_{1}$ and $\chi_{2}$ are variables with Chi-squared distribution with one 
degree of freedom,  $\alpha_1 = L_{j}(d) - 2L_{j}^{*}(d)$, 
$\alpha_2 = L_{j}(d)+ 2L_{j}^{*}(d)$,
\begin{equation}
L_{j}(d)= \lim_{n\rightarrow\infty} \mathbb{E}\left\{\frac{I_{x}(\omega_j)}{f_{X}(\omega_j)}\right\}
=\frac{2}{\pi}\int^{\infty}_{-\infty}\frac{\sin^{2}(\omega/2)}{(2\pi j -\omega)^{2}}\left|
\frac{\omega}{2\pi j}\right|^{-2d}d\omega
\end{equation}
and
\begin{equation}
L_{j}^{*}(d) =\frac{1}{\pi}\int^{\infty}_{-\infty}\frac{\sin^{2}(\omega/2)}
{(2\pi j -\omega)(2\pi j +\omega)}\left| \frac{\omega}{2\pi j}\right|^{-2d}d\omega.
\end{equation}

Let $\{Z_t\}_{t\in\mathbb{Z}}$ be a process contaminated by additive
outliers, which is described by
\begin{align}\label{eq:ao}
Z_t=X_t+\sum_{j=1}^m\varpi_j Y_{j,t},
\end{align}
where $m$ is the maximum number of outliers; the unknown parameter $\omega_j$
represents the magnitude of the $j$th outlier, and
$Y_{j,t}\,(\equiv Y_j)$ is a random variable (\textit{r.v.}) with probability
distribution $\Pr\left(Y_j=-1\right)=\Pr\left(Y_j=1\right)=\frac{p_j}{2}$ and
$\Pr\left(Y_j=0\right)=1-p_j$, where $\mathbb{E}[Y_j]=0$ and
$\mathbb{E}[Y_j^2]=\mathrm{var}(Y_j)=p_j$. Model \ref{eq:ao} is based on the
parametric models proposed by \citeasnoun{fox:1972}. $Y_j$ is the product of 
$Bernoulli(p_j)$ and \textit{Rademacher} random variables; the latter equals $1$ or 
$-1$, both with probability $\frac12$. $X_t$ and $Y_j$ are independent random variables.

Some results related to the effects of outliers on the spectral density and on the 
autocorrelation functions of $\{Z_t\}_{t\in\mathbb{Z}}$ are presented as follows.
\begin{prop}\label{prop:espectro}
Suppose that $\{Z_t\}_{t\in \mathbb{Z}}$ follows Model \ref{eq:ao}.
\begin{enumerate}
\item[$i.$] The autocovariance function (ACOVF) of $\{Z_t\}_{t\in\mathbb{Z}}$ is given 
by
\begin{align*}
\gamma_z(h)=\gamma_X(h)+\delta'(h)\sum\limits_{j=1}^m\varpi_j^2p_j,
\end{align*}
\noindent where
$\gamma_X(h)=\mathbb{E}[X_tX_{t+h}]-\mathbb{E}[X_t]\mathbb{E}[X_{t+h}]$,
$\delta'(h)=
\begin{cases}
1, & \text{when } h=0,\\
0, & \text{otherwise}.
\end{cases}$ with $h\in\mathbb{Z}$.
\item[$ii.$] The spectral density function of $\{Z_t\}$ is given by
\begin{align*}
f_Z(\omega)=f_X(\omega)+\frac{1}{2\pi}\sum_{j=1}^m\varpi_j^2p_j, \quad
\omega\in(-\pi,\pi],
\end{align*}
where $f_X(\omega)=\dfrac{1}{2\pi}\sum\limits_{h=-\infty}^\infty
\gamma_X(h)e^{-ih\omega}$.
\end{enumerate}
\end{prop}
\noindent
Proposition \ref{prop:espectro} states that
$\gamma_z(h)$, for $ h=0$, depends on $\mathrm{var} (Y_j)$. $\gamma_Z(0)$ increases with
$\mathrm{var}(Y_j)$ (see the proof in \citeasnoun{fajardo:reisen:cribari:2009}). This
relation between $R_Z(0)$ and $\mathrm{var}(Y_j)$ will certainly affect the model
parameter estimates  because it reduces the magnitude of the autocorrelations and
introduces loss of information on the pattern of serial correlation (see also 
\citename{chan:1995} \citeyear{chan:1992,chan:1995})

The  spectral form of  $\{Z_t\}_{t\in\mathbb{Z}}$ (Model \ref{eq:ao}) when 
$\{X_t\}_{t\in\mathbb{Z}}$ follows an ARFIMA($p,d,q$) model is given in the next
lemma.
\begin{lema}\label{lema:lema1}
Let $\{X_t\}_{t\in\mathbb{Z}}$ be a stationary and invertible ARFIMA($p,d,q$)
process. Also, let $\{Z_t\}_{t\in\mathbb{Z}}$ be such that
$Z_t=X_t+\sum_{j=1}^m\varpi_jY_j$, where $m$ is the maximum number of outliers, the
unknown parameter $\varpi_j$ is the magnitude of the $j$th outlier and $Y_j$ is a
r.v. with probability distribution
$\Pr\left(Y_j=-1\right)=\Pr\left(Y_j=1\right)=\frac{p_j}{2}$ and
$\Pr\left(Y_j=0\right)=1-p_j$. The spectral density of $\{Z_t\}_{t\in\mathbb{Z}}$
is
\begin{align*}
f_Z(\omega)&=\frac{\sigma_\epsilon^2}{2\pi}\frac{|\Theta(e^{-i\omega})|^2}{|\Phi(e^{-i\omega})|^2}
\left\{2\sin\left(\frac{\omega}{2}\right)\right\}^{-2d}+\frac{1}{2\pi}\sum_{j=1}^m\varpi_j^2p_j.
\end{align*}
\end{lema}
\noindent The proof of Lemma \ref{lema:lema1} follows directly
from Proposition \ref{prop:espectro}.

The effects of an outlier on the sample autocovariance function and on the periodogram 
are given below.
\begin{prop}\label{prop:sesgo}
Let $z_1,z_2,\ldots,z_n$ be generated from Model \ref{eq:ao} with one outlier, and let 
the outlier occur at time $t=T$ with $h<T<n-h$. It follows that:
\begin{enumerate}
\item[$i.$] The sample ACOVF is given by
\begin{align}\label{eq:cov_cont}
\widehat{\gamma}_z(h)&=\widehat{\gamma}_x(h)\pm\frac{\varpi}{n}(x_{_{T-h}}+x_{_{T+h}}
-2\bar{x})+\frac{\omega^2}{n}\delta'(h)+ o_p(n^{-1}),
\end{align}
where
$\widehat{\gamma}_x(h)=\dfrac1n\sum\limits_{t=1}^{n-h}(x_t-\bar{x})(x_{t+h}-\bar{x})$.

\item[$ii.$] The periodogram is given by
\begin{align*}
I_z(\omega)&= I_x(\omega)+\Delta(\varpi), \quad \omega\in(-\pi,\pi],
\end{align*}
\noindent
where
$I_x(\omega)=\dfrac{1}{2\pi}\sum\limits_{h=-(n-1)}^{n-1}\widehat{\gamma}_x(h)e^{-ih\omega},$
and
\end{enumerate}
\begin{align*}
\Delta(\varpi)&=\frac{\varpi^2}{2\pi n}\pm\frac{\varpi}{\pi
n}\left\{(x_{_T}-\bar{x})+\sum_{h=1}^{n-1}(x_{_{T-h}}+x_{_{T+h}}-2\bar{x})\cos(h\omega)\right\}+o_p(n^{-1}).
\end{align*}
\end{prop}
These results show that outliers may substantially affect the inference performed on 
stationary models by revealing  that there is information loss in the serial correlation
dynamics of the process, which is translated into the parameter estimation process.

\section{ The autocovariance and spectral density robust functions}\label{ssec:estt}
\subsection{The autovariance function}
\citeasnoun{ma:genton:2000} proposed a scale covariance estimator which is
based on $Q_n(\cdot)$, defined in the sequel, and on the following covariance identity
\begin{align}\label{eq:cov_escal}
\textrm{cov}(X,Y)=\frac{1}{4ab}[\textrm{var}(aX+bY)-\textrm{var}(aX-bY)],
\end{align}
where $X$ and $Y$ are random variables, $a=\frac{1}{\sqrt{\textrm{var}(X)}}$
and $b=\frac{1}{\sqrt{\textrm{var}(Y)}}$ \cite{huber:2004}.

\citeasnoun{rousseeuw:croux:1993} proposed a robust scale estimator function
$Q_n(\cdot)$ which is based on the $\tau$th order statistic of
$\binom{n}{2}$ distances $\{|\eta_j-\eta_k|,j<k\}$, and can be written as
\begin{align}\label{eq:Q}
Q_n(\eta)=c\times\{|\eta_j-\eta_k|;j<k\}_{(\tau)},
\end{align}
where $\eta=(\eta_1,\eta_2,\ldots,\eta_n)'$, $c$ is a constant used to guarantee
consistency ($c=2.2191$ for the normal distribution) and
$\tau=\left\lfloor\frac{\binom{n}{2}+2}{4}\right\rfloor+1$.

Based on identity (\ref{eq:cov_escal}) and on $Q_n(\cdot)$,
\citeasnoun{ma:genton:2000} proposed a highly robust estimator for the
ACOVF:
\begin{align}\label{eq:acovr}
\widehat{\gamma}_Q(h)=\frac14\left[Q_{n-h}^2({\bf u}+{\bf v})-Q_{n-h}^2({\bf u}-{\bf
v})\right],
\end{align}
\noindent
where ${\bf u}$ and ${\bf v}$ are vectors containing the initial $n-h$ and the
final $n-h$ observations, respectively. The robust estimator for the autocorrelation
function (ACF) is
\begin{align*}
\widehat{\rho}_Q(h)=\frac{Q_{n-h}^2({\bf u}+{\bf v})-Q_{n-h}^2({\bf u}-{\bf
v})}{Q_{n-h}^2({\bf u}+{\bf v})+Q_{n-h}^2({\bf u}-{\bf v})}.
\end{align*}
\noindent It can be shown that $|\widehat{\rho}_Q(h)|\le 1$ for all $h$.

\subsubsection{Influence Function and Breakdown Point.}
Influence Function (IF) is an important tool to understand the effect of the 
contamination of an outlier in any estimator. To define IF supposes that the empirical
cumulative distribution function $F_n$ of $x_1,..x_n$, adequately normalized, converges.
Following \citeasnoun{huber:2004}, the influence function $x \rightarrow IF(x,T,F)$ is 
defined for a functional $T$ at a distribution $F$ and at point $x$ as the limit
\begin{equation*}
IF(x,T,F) = \lim_{\varepsilon\to 0+}\varepsilon^{-1}\{T(F+\varepsilon(\delta_x-F))-T(F)\}\; ,
\end{equation*}
where $\delta_x$ is the Dirac distribution  at $x$.

Breakdown Point (BP) indicates the largest proportion of outliers that the data may
contain such that the estimator still gives some information about the distribution
of the outlier-free data (\citeasnoun{maronna:martin:yohai:2006}). 
\citeasnoun{rousseeuw:croux:1993} showed that the asymptotic BP of $Q_n(\cdot)$ is 
$50\%$, which means that the data can be contaminated by up to half of the observations
with outliers and $Q_n(\cdot)$ will still yield sensible estimates.

The classical notion of sample BP of a scale estimator $S_n(\cdot)$ is given in
Definition \ref{def:BP}.
\begin{defin} \label{def:BP}%
Let $\eta=(\eta_1, \eta_2,\ldots, \eta_n)'$ be a sample of size $n$. Let
$\widetilde{\bf \eta}$ be obtained by replacing any $m$ observations of ${\bf
\eta}$ by arbitrary values. The sample breakdown point of a scale estimator
$S_n({\bf \eta})$ is given by
\begin{align*}
\varepsilon_n^*(S_n({\bf \eta}))&=\max\left\{\frac{m}{n}:\sup_{\widetilde{\bf
\eta}}S_n(\widetilde{\bf \eta})<\infty \text{ and } \inf_{\widetilde{\bf
\eta}}S_n(\widetilde{\bf \eta})>0\right\}.
\end{align*}
\end{defin}
\noindent The above BP definition holds for a scale estimator function of a
time invariant random sample.
As noted by \citeasnoun{ma:genton:2000}, in time series, the estimators are based on
differences between observations apart by various time lag distances and usually
have a BP with respect to these differences. Then,  the time location of the
outlier becomes important (see also, for example, \citeasnoun{ledolter:1989}).
Therefore, the authors introduced the following definition of a temporal sample
breakdown point of an autocovariance estimator $\hat\gamma_{\bf \eta}(h)$ based on
(\ref{eq:cov_escal}).

\begin{defin} \label{def:BPcov}%
Let ${\bf \eta}=(\eta_1, \eta_2,\ldots, \eta_n)'$ be a sample of size $n$ and let
$\widetilde{\bf \eta}$ be obtained by replacing any $m$ observations of ${\bf
\eta}$ by arbitrary values. Denote by $\mathbb{I}_m$ a subset of size $m$ of
$\{1,2,\ldots,n\}$. The temporal sample breakdown point of an autocovariance
estimator $\hat\gamma_{\bf \eta}(h)$ is given by
\begin{align*}
\varepsilon_n^{temp}(\widehat\gamma_{\bf
\eta}(h))&=\max\left\{\frac{m}{n}:\sup_{\mathbb{I}_m}\sup_{\widetilde{\bf
\eta}}S_{n-h}(\widetilde{\bf u}+\widetilde{\bf v})<\infty,
\inf_{\mathbb{I}_m}\inf_{\widetilde{\bf \eta}}S_{n-h}(\widetilde{\bf
u}+\widetilde{\bf v})>0, \right.\\
&\left.\sup_{\mathbb{I}_m}\sup_{\widetilde{\bf \eta}}S_{n-h}(\widetilde{\bf
u}-\widetilde{\bf v})<\infty \text{ and } \inf_{\mathbb{I}_m}\inf_{\widetilde{\bf
\eta}}S_{n-h}(\widetilde{\bf u}-\widetilde{\bf v})>0 \right\},
\end{align*}
\noindent where $\widetilde{\bf u}$ and $\widetilde{\bf v}$ are derived from
$\widetilde{\bf \eta}$ as in (\ref{eq:acovr}).
\end{defin}
\begin{note}
The relation between the classical sample and the temporal sample breakdown points
can be expressed by the following inequality (\citeasnoun{ma:genton:2000}):
\begin{align*}
    \frac{n-h}{2n}\varepsilon_n^*(\widehat\gamma_{\bf
    \eta}(h))\le\varepsilon_n^{temp}(\widehat\gamma_{\bf
    \eta}(h))\le\frac12\varepsilon_n^*(\widehat\gamma_{\bf \eta}(h)).
\end{align*}
\end{note}
\noindent It then follows that since the sample breakdown point of
the classical autocovariance estimator is zero, the temporal breakdown point of this
estimator is also zero. This means that only one single outlier is enough to
`break' the estimator.

\citeasnoun{ma:genton:2000} showed that the maximum  temporal breakdown point of the 
highly robust autocovariance estimator is 25\%, which is the highest possible breakdown
point for an autocovariance estimator.

Results of the asymptotic properties of the robust autocovariance function for
a Gaussian ARFIMA model are summarized as follows 
(see \citeasnoun{levy:boistard:moulines:taqqu:reisen:2011a}).

\subsubsection{Short-memory case}
Let $\{X_{t}\}_{t\in\mathbb{Z}}$ be a stationary mean-zero Gaussian process given by 
Model \ref{eq:arfima} with $d=0$, that is, the autocovariance function 
($\gamma(h)=E(X_{1}X_{h+1}))$ of $\{X_{t}\}_{t\in\mathbb{Z}}$ satisfies
\begin{equation*}
\sum_{h\geq 1}|\gamma(h)|<\infty.
\end{equation*}

The following theorems present the asymptotic behavior of the robust autocovariance 
estimator.
\begin{thm}\label{theo:gamma_short} 
Let $h$ be a non-negative integer. Under the assumption that the autocovariances are 
absolutely summable, the autocovariance estimator 
$\widehat{\gamma}_Q(h,\chunk{X}{1}{n},\Phi)$ satisfies the following Central Limit 
Theorem:
\begin{equation*}
\sqrt{n}\left(\widehat{\gamma}_Q(h,\chunk{X}{1}{n},\Phi)-\gamma(h)\right)
\stackrel{d}{\longrightarrow}\mathcal{N}(0,{\check{\sigma}}^2_h),
\end{equation*}
\noindent where
\begin{equation}\label{e:lvar2}
\check{\sigma}^2(h)=E[\psi^2(X_1,X_{1+h})]+2\sum_{k\geq 1} E[\psi(X_1,X_{1+h}) \psi(X_{k+1},X_{k+1+h})]\;
\end{equation}
where $\psi$ is a function of $\gamma(h)$ and of IF (see, Theorem 4 in 
\citeasnoun{levy:boistard:moulines:taqqu:reisen:2011a}).
\end{thm}

\subsubsection{Long-memory case}
Now, let $d\neq 0$ in  Model \ref{eq:arfima} and let $D=1-2d$. The ACF behaves like
\begin{equation*}
\gamma(h)=h^{-D} L(h),\ 0<D<1\; ,
\end{equation*}
where $L$ is slowly varying at infinity and is positive for large $h$.
Note that, for positive $d$, as previously stated, the ACF of the process
is not absolutely summable.

\begin{thm}\label{theo:gamma_long}
Let $h$ be a non negative integer. Then, $\widehat{\gamma}_Q(h,\chunk{X}{1}{n},\Phi)$ 
satisfies the following limit theorems as $n$ tends to infinity.
\begin{itemize}
\item If $D>1/2$,
\begin{equation*}
\sqrt{n}\left(\widehat{\gamma}_Q(h,\chunk{X}{1}{n},\Phi)-\gamma(h)\right)
\stackrel{d}{\longrightarrow}\mathcal{N}(0,{\check{\sigma}}^2(h))\; ,
\end{equation*}
where
\begin{equation*}
\check{\sigma}^2(h)=\mathbb{E}[\psi^2(X_1,X_{1+h})]+2\sum_{k\geq 1} 
\mathbb{E}[\psi(X_1,X_{1+h}) \psi(X_{k+1},X_{k+1+h})]\; ,
\end{equation*}
where $\psi$ is a function of $\gamma(h)$ and of IF (see, Theorems 4 and 5 in 
\citeasnoun{levy:boistard:moulines:taqqu:reisen:2011a}).
\item If $D<1/2$,
\begin{equation*}
\beta(D)\frac{n^{D}}{\widetilde{L}(n)}\left(\widehat{\gamma}_Q(h,\chunk{X}{1}{n},\Phi)-\gamma(h)\right)
\stackrel{d}{\longrightarrow}
\frac{\gamma(0)+\gamma(h)}{2}(Z_{2,D}(1)-Z^2_{1,D}(1))
\end{equation*}
where  $\beta(D)=\emph{B}((1-D)/2,D)$, $\emph{B}$ denotes the Beta function, the 
processes $Z_{1,D}(\cdot)$ and $Z_{2,D}(\cdot)$ are defined by Equations $53$ and $54$,
respectively, in \citeasnoun{levy:boistard:moulines:taqqu:reisen:2011a}, and
\begin{equation} \label{eq:expression-tildeL}
\widetilde{L}(n)=2L(n)+L(n+h)(1+h/n)^{-D}+L(n-h)(1-h/n)^{-D}\;.
\end{equation}
\end{itemize}
\end{thm}

\begin{note}
For  Model \ref{eq:arfima} with $ 1/4 < d < 1/2$, the robust autocovariance estimator 
$\widehat{\gamma}_Q(h,\chunk{X}{1}{n},\Phi)$ has the same asymptotic behavior as the 
classical autocovariance estimator $\widehat{\gamma}_x(h)$.
\end{note}

Theories related to the use of the robust ACF function  to obtain an spectral estimate 
are still opened questions. However, this was first empirically investigated by 
\citeasnoun{fajardo:reisen:cribari:2009}.
The authors considered a robust estimator of the spectral density based on the robust 
ACF function when the time series follows an ARFIMA Model. Their estimation method is 
discussed in the next sub-section.


\subsection{The sample spectral function}
The results discussed in the previous sections and the spectral representation of a 
stationary process justify the use of the robust ACF function in the calculus of an 
estimator of a spectral density.

As previously stated, for the stationary process $\{X_t\}_{t\in\mathbb{Z}}$, the 
spectral density is a real-valued function of the Fourier transform of the 
autocovariance function, that is,
\begin{equation}\label{EspcRepre}
f_X(\omega)=\dfrac{1}{2\pi}\sum\limits_{h=-\infty}^\infty
\gamma_X(h)e^{-ih\omega}
\end{equation}
where $\gamma_X(\cdot)$ is the autocovariance of the process.

Equation \ref{EspcRepre} suggests to replace  $\gamma_X(\cdot)$ by its estimate to 
obtain an estimate of $f_X(\omega)$. The periodogram function is the classical tool 
to estimate the spectral function. Other variants of the periodogram are called smoothed
window periodogram (see, e.g., \citeasnoun{priestley:1981}).
In the same direction, \citeasnoun{fajardo:reisen:cribari:2009} suggested to use the 
robust autocovariance function as an estimator of the classical ACF to obtain a robust
spectral function. Although the theoretical justification of this estimator is still an
opened question, the authors have empirically shown that the robust spectral estimator 
can be an alternative method to estimate a time series with outliers. A robust spectral
estimator is
\begin{align}\label{eq:perr}
I_Q(\omega)&=\frac{1}{2\pi}\sum_{|h|<n}\kappa(h)\widehat{\gamma}_Q(h)\cos(h\omega),
\end{align}
\noindent where $\widehat{\gamma}_Q(h)$ is the sample autocovariance function given in
Equation \ref{eq:acovr} and $\kappa(h)$ is defined as
\begin{align*}
\kappa(h)=
\begin{cases}
1, & |h|\le M,\\
0, & |h|> M.
\end{cases}
\end{align*}
\noindent $\kappa(h)$ is a particular case of  the \textit{lag window} functions
used in classical spectral theory to obtain a consistent spectral estimator,
and $M$ is the truncation point which is a function of $n$, say $M = G(n)$,  where
$G(n)$ must satisfy $G(n) \rightarrow \infty $, $n \rightarrow \infty $, with
$\frac{G(n)}{n} \rightarrow \ 0$. $G(n)$ is usually chosen to
be $G(n)=n^\beta$, where $0<\beta<1$ (see, e.g. 
\citeasnoun[pp.\ 433--437]{priestley:1981}). Note that, equivalently to the classical 
spectral estimation theories, other different \textit{lag window} functions can be used
to obtain a robust spectral estimator.

Since Equation \ref{eq:perr} does not have the same finite-sample properties as the 
periodogram, it  is defined here as  \textit{robust truncated pseudo-periodogram}. For 
large $h$, the numbers of observations in the calculus of  $\widehat{\gamma}_Q(h)$ are 
very small and, consequently,  this function  becomes very unstable. Then, to avoid 
these undesirable covariance estimates in the calculus of the estimator given in 
Equation \ref{eq:perr}  justify the use of a   truncation point $M$ in the calculus of
this sample function (see \citeasnoun{fajardo:reisen:cribari:2009}). The authors 
suggested $M$ that satisfies

\begin{align*}
M\le h'=\min\left\{0<h<n:\varepsilon_n^{temp}\left(\widehat{\gamma}_Q(h)\right)\le\frac{m}{n}\right\}-1.
\end{align*}

%
\section{Semiparametric estimation methods of $d$ and empirical studies}
The semiparametric estimation procedure based on the OLS estimator proposed by 
\citeasnoun{geweke:porter-hudak:1983}(GPH) is considered. Since the GPH estimator is 
well-discussed in the literature, this method and its asymptotic statistical properties
are briefly summarized as follows.

For a single realization $x_1,\ldots,x_n$ of $\{X_t\}_{t\in \mathbb{Z}}$, the GPH
estimate of $d$ is obtained from the regression equation
\begin{equation} \label{eqeREGS1}
\log I_x(\omega_j)= a_{0}-2d\log{[2 \sin(\omega_j/2)]}+\xi_j, j=1,\ldots,m'
\end{equation}
\noindent where $\omega_j$ is the Fourier frequency at $j$, $m'$ is the bandwidth in the
regression equation which has to satisfy 
$m' \rightarrow \infty $, $n \rightarrow \infty$, with $\frac{m'}{n} \rightarrow \ 0$ 
and $ \frac{m'\log (m')}{n} \rightarrow \ 0$,  $a_{0}$= $\log f_\eta(0) $ + $\log
\frac{f_\eta(\omega_j)}{f_\eta(0)}$ + C, 
$\xi_j = \log\frac{I_x(\omega_j)}{f_X(\omega_j)}- C$ and C = $\varphi(1)$ ($\varphi(.)$
is the digamma function).

The GPH estimate of $d$ is given by
\begin{equation} \label{eqeReg}
     d_{GPH} = (-0.5)\frac{\sum_{j=1}^{m'}(v_{j} - \bar{v})\log I_x(\omega_j) }{S_{vv}}
\end{equation}
where $S_{vv} =\sum_{j=1}^{m'}(v_j -\overline{v})^2$, $v_{j}= \log \left\{4\sin^{2}(\omega_{j}/2)\right\}$.

Under some conditions, \citeasnoun{hurvich:deo:brodsky:1998} proved that the 
GPH-estimator is consistent for the memory parameter and asymptotically normal for 
Gaussian time series processes. The authors established that the optimal $m'$ in 
Equations \ref{eqeREGS1} and \ref{eqeReg} is of order $o(n^{4/5})$ and 
$(m')^{1/2}(d_{GPH}-d)$ $\overset{d}{\longrightarrow} N(0,\frac{\pi^2}{24})$.

To obtain a robust estimator of $d$, \citeasnoun{fajardo:reisen:cribari:2009} proposed 
to replace  in Equation \ref{eqeREGS1} the $\log I_x(\omega_j)$ by $\log{I_Q(\omega_j)}$
which gives the following OLS regression estimator

\begin{align}\label{eq:GPHrob}
d_{GPHR}=-(0.5)\frac{\sum_{j=1}^{m^{'}}(\upsilon_j-\bar{\upsilon})\log{I_Q(\omega_j)}}{S_{vv}},
\end{align}
\noindent where
$S_{vv}$, $m'$ are defined as before and $I_Q(\omega)$ is the function given in 
Equation \ref{eq:perr}. As previously mentioned, the asymptotical properties of 
$d_{GPHR}$ still remains to be established. However, based on the following empirical 
investigation, the robust method seems to be a reasonable robust alternative method to 
estimate long-memory time series in the presence of additive outliers.

\subsection{Numerical evaluation using the ARFIMA$(0,d,0)$ model}
The finite series were simulated from zero-mean ARFIMA models (Eq. \ref{eq:arfima}) 
with $\{\epsilon_t\}_{t\in \mathbb{Z}}$, $t=1,...,n$, i.i.d.  $N(0,1)$. The models, 
parameters, sample sizes and  empirical results are displayed in the following tables.
The empirical mean, standard deviation (s.d.), bias and mean squared error (MSE) were 
obtained as a mean of 10.000 replications. The contaminated data were generated from 
Model \ref{eq:ao} with $m=1$, $p=0.05$ for magnitude $\varpi=10$  and bandwidth values
for $d_{GPH}$ and $d_{GPHR}$ were computed for $\alpha=0.7$ and truncation point 
$M = n^\beta$, $\beta=0.7$. In the tables $d_{GPH_c}$ and $d_{GPHR_c}$ mean the 
estimates of $d$ when the series has outliers. The simulations were carried out using 
the \texttt{Ox} matrix programming language (see \texttt{http://www.doornik.com}).
The empirical study is divided into the following  model properties: stationary and 
non-stationary processes.

\subsubsection{Stationary model}
Table \ref{tab:d03} displays  results for $d=0.3, 0.45$ and $\alpha=\beta=0.7$.
From this it can be seen that when the series does not contain outliers, both
estimators present similar behavior in the estimation of $d$,  which is not a surprise 
result. However, the introduction of outliers in the series dramatically changes the 
performance of the classical estimator (GPH), in particular, it significantly 
underestimates the true parameter. On the other hand, in this scenario, the robust 
method (GPHR) seems to be not sensitive to outliers. Other cases were also simulated 
such as ARFIMA models with AR and MA parts and different  values of $p$ and $\varpi$.
All cases indicated similar conclusions to the one given in Table \ref{tab:d03}. These 
are available upon request. Table 2 gives the estimates of $d$ when different 
lag-windows are used to compute the robust periodogram estimator.
The lag-windows are Parzen (P), Tukey-Hamming(TH) and Bartlett (B) and the fractional 
estimators were computed with the same bandwidths as in the previous case and the 
results are in Table \ref{tab:diflagwindow}. The choice of the lag-window does not 
appear to be too important in the estimation of $d$ since the estimates obtained from 
different lag-windows are, in general, numerically very close to each other, that is, 
the estimates are not too sensitive to the choice of the lag-window.  These lag-windows
yield similarly accurate estimates compared to the one given in Equation \ref{eq:perr}.

\begin{table}[!htb]
{\footnotesize
\begin{center}\caption{\footnotesize{Simulation results:
ARFIMA$(0,d,0)$ model with $\alpha=\beta=0.7$ and $\varpi=0,10$.}}
\begin{tabular}{c|c|l|rr|rr}\hline\hline
$d$    & $n$ &          & ${d}_{GPH}$   & ${d}_{GPH_c}$ & ${d}_{GPHR}$ &
${d}_{GPHR_c}$\\ \hline\hline
       & 100 &  mean & $ 0.2988$  & $ 0.1134$  & $ 0.2584$  & $ 0.2449$\\
       &     &  s.d.      & $ 0.1735$  & $ 0.1619$  & $ 0.1558$  & $ 0.1556$ \\
       &     &  bias  & $-0.0012$  & $-0.1866$  & $-0.0416$  & $-0.0551$\\
       &     &  MSE     & $ 0.0301$  & $ 0.0610$  & $ 0.0260$  & $ 0.0272$\\
\cline{2-7}
       & 300 &  mean & $0.3062$   & $ 0.1007$  & $ 0.2907$  & $ 0.2837$\\
$0.30$ &     &  s.d.      & $0.1005$   & $ 0.0978$  & $ 0.0926$  & $ 0.0960$ \\
       &     &  bias  & $0.0062$   & $-0.1993$  & $-0.0093$  & $-0.0163$\\
       &     &  MSE     & $0.0101$   & $ 0.0493$  & $ 0.0087$  & $ 0.0095$\\
\cline{2-7}
       & 800 &  mean & $0.3003$  & $ 0.1184$  & $ 0.2949$  & $ 0.2869$\\
       &     &  s.d.      & $0.0679$  & $ 0.0715$  & $ 0.0573$  & $ 0.0610$ \\
       &     &  bias  & $0.0003$  & $-0.1816$  & $-0.0051$  & $-0.0131$\\
       &     &  MSE    & $0.0046$  & $ 0.0381$  & $ 0.0033$  & $ 0.0039$\\ \hline
       & 100 &  mean & $0.4561$  & $ 0.1923$  & $ 0.3975$  & $ 0.3778$\\
       &     &  s.d.      & $0.1722$  & $ 0.1727$  & $ 0.1506$  & $ 0.1433$ \\
       &     &  bias  & $0.0061$  & $-0.2577$  & $-0.0525$  & $-0.0722$\\
       &     &  MSE     & $0.0297$  & $ 0.0962$  & $ 0.0254$  & $ 0.0258$\\
\cline{2-7}
       & 300 &  mean   & $0.4594$  & $ 0.2015$  & $ 0.4329$  & $ 0.4233$\\
$0.45$ &     &  s.d.        & $0.0986$  & $ 0.0976$  & $ 0.1041$  & $ 0.1013$ \\
       &     &  bias    & $0.0094$  & $-0.2485$  & $-0.0171$  & $-0.0267$\\
       &     &  MSE       & $0.0098$  & $ 0.0713$  & $ 0.0111$  & $ 0.0110$\\
\cline{2-7}
       & 800 &  mean   & $0.4620$  & $ 0.2306$  & $ 0.4457$  & $ 0.4349$\\
       &     &  s.d.        & $0.0688$  & $ 0.0809$  & $ 0.0562$  & $ 0.0576$ \\
       &     &  bias    & $0.0121$  & $-0.2194$  & $-0.0043$  & $-0.0151$\\
       &     &  MSE       & $0.0049$  & $ 0.0547$  & $ 0.0032$  & $ 0.0035$\\
       \hline\hline
\end{tabular}\label{tab:d03}
\end{center}}
\end{table}

\begin{table}[!ht]
{\footnotesize
\begin{center}\caption{\footnotesize{Empirical results of  $d$'s estimators in
ARFIMA$(0,d,0)$ model using different lag-windows.}}
\begin{tabular}{c|c|l|r|r|r}\hline\hline
\multicolumn{6}{c}{uncontaminated series}\\ \hline\hline
Parameter &$n$&&$d_{P}$&$d_{TH}$&$d_{B}$\\ \hline\hline
	&100&mean &$ 0.2699$&$ 0.2602$&$ 0.2459$\\
	&   &s.d. &$ 0.1497$&$ 0.1575$&$ 0.1444$\\
	&   &bias &$-0.0301$&$-0.0398$&$-0.0541$\\
	&   &MSE  &$ 0.0233$&$ 0.0264$&$ 0.0238$\\
\cline{2-6}
        &300&mean&$ 0.2880$&$ 0.2833$&$ 0.2857$\\
$d=0.3$ &   &s.d.&$ 0.1050$&$ 0.1037$&$ 0.0976$\\
	&   &bias&$-0.0119$&$-0.0167$&$-0.0143$\\
	&   &MSE &$ 0.0112$&$ 0.0110$&$ 0.0097$\\
\cline{2-6}
	&800&mean&$ 0.2985$&$ 0.2966$&$ 0.3001$\\
	&   &s.d.&$ 0.0554$&$ 0.0584$&$ 0.0561$\\
	&   &bias&$-0.0015$&$-0.0034$&$ 0.0001$\\
	&   &MSE &$ 0.0031$&$ 0.0034$&$ 0.0031$\\ \hline\hline
\multicolumn{6}{c}{contaminated series}\\ \hline\hline
Parameter &$n$&&$d_{P}$&$d_{TH}$&$d_{B}$\\ \hline\hline
	&100&mean&$ 0.2504$&$ 0.2446$&$ 0.2419$\\
	&   &s.d.&$ 0.1552$&$ 0.1482$&$ 0.1405$\\
	&   &bias&$-0.0496$&$-0.0554$&$-0.0581$\\
	&   &MSE &$ 0.0266$&$ 0.0250$&$ 0.0231$\\
\cline{2-6}
        &300&mean&$ 0.2806$&$ 0.2729$&$ 0.2796$\\
$d=0.3$ &   &s.d.&$ 0.1028$&$ 0.0925$&$ 0.0964$\\
	&   &bias&$-0.0194$&$-0.0271$&$-0.0204$\\
	&   &MSE &$ 0.0109$&$ 0.0093$&$ 0.0097$\\
\cline{2-6}
	&800&mean&$ 0.2934$&$ 0.2889$&$ 0.2928$\\
	&   &s.d.&$ 0.0578$&$ 0.0606$&$ 0.0553$\\
	&   &bias&$-0.0066$&$-0.0111$&$-0.0072$\\
	&   &MSE &$ 0.0034$&$ 0.0038$&$ 0.0031$\\
\hline\hline
\end{tabular}\label{tab:diflagwindow}
\end{center}}
\end{table}

\subsubsection{Non-stationary model}
As is well-known, the GPH estimator has been widely used even in the case when the 
ARFIMA model has  $d$ in $(0.5,1.0]$ (see, for example, \citeasnoun{franco:reisen:2007},
\citeasnoun{hurvich:ray:1995}, \citeasnoun{olbermann:lopes:reisen:2006}, 
\citeasnoun{phillips:2007} among others).

Based on the theory discussed in the previous sections, the robust method can not be 
applied in a non-stationary time series. However, it may be interesting to verify if 
GPHR estimator is invariant to the first difference, i.e. the estimative of memory 
parameter based on the original data is equal to one plus the estimated $d$ based on the
differenced data.

Now, let  Model \ref{eq:arfima} be defined with  parameter $d^*=d + \kappa$,
where $d \in (-0.5,0.5)$, $\kappa>0$, $\kappa\in\mathbb{Z}$. Then,
Model \ref{eq:arfima}, with zero-mean, becomes

\begin{equation}\label{eq:nonstarfima}
X_{t} =(1-{B})^{-d^*}\eta_t, \quad t\in \mathbb{Z}.
\end{equation}

Process given in Equation \ref{eq:nonstarfima} is non-stationary when $d^*\geq 0.5$;
however, it is still persistent. For $d^*\in [0.5,1.0)$ it is
level-reverting in the sense that there is no long-run
impact of an innovation on the value of the process. The level-reversion
property no longer holds  when $d^*\geq 1$. Note that when  $d^*= 1$ the process is a 
random walk.

From  Model \ref{eq:nonstarfima} with $\kappa=1$ and $p=q=0$
\begin{equation*}
W_{t}=(1-{ B}){X}_t,  t\in\mathbb{Z},
\end{equation*}
is an $ARFIMA(0,d,0)$ process.  Let ${\hat d^*}$ be the estimator of $d^*$ and $\hat d$
the fractional estimator obtained from the differenced data. The main goal is to verify
the equality ${\hat d^*}= {\hat d} + 1$ for uncontaminated and contaminated series.
Based on the same simulation procedure previously described, series from 
Model \ref{eq:nonstarfima} were generated and some cases are displayed in 
Table \ref{tab:diff} (other cases are available upon request). Similar conclusions to 
the previous study are observed. Both estimators present equivalent performance when 
they are applied in the first difference of uncontaminated series. This suggests that 
both can be used in practical situations when dealing with  non-stationary data. 
However, since the first difference does not eliminate the effect of an outlier, the 
estimates  clearly indicate that caution has to be exercised when there is  suspicion 
of outliers in the data. The GPH estimator  presents  poor performance in terms of bias
(high positive bias) and $MSE$. In  contrast to the GPH estimator,   the GPHR method 
seems to be invariant to the first difference of  non-stationary time series with 
outliers. This empirical study suggests that, in practical situations when dealing with
non-stationary data with outliers, one solution is to apply the first difference in the
series and then to estimate $d$ with the  robust estimator discussed in this paper.

\begin{table}[!htb]
{\footnotesize
\begin{center}\caption{\footnotesize{Empirical results: ARFIMA$(0,d,0)$ model with differenced data
and $\omega=0,10$.}}
\begin{tabular}{c|c|l|rr|rr}\hline\hline
Parameter &$n$&&${d}_{GPH}$   & ${d}_{GPH_c}$ & ${d}_{GPHR}$ &
${d}_{GPHR_c}$\\
\hline\hline
\cline{2-7}
            & 300 &  mean & $-0.2141$  & $-0.5066$   & $-0.1906$    & $-0.2211$
\\
$d_X=0.8, d_W=-0.2$      &     &  bias & $ 0.0141$  & $ 0.3066$   & $-0.0094$
& $ 0.0211$ \\
            &     &  s.d  & $ 0.1076$  & $ 0.1469$   & $ 0.1127$    & $ 0.1421$
\\
            &     &  MSE  & $ 0.0118$  & $ 0.1155$   & $ 0.0128$    & $ 0.0206$
\\
\cline{2-7}
            & 800 &  mean & $-0.1906$  & $-0.4283$   & $-0.2062$    & $-0.2250$
\\
            &     &  bias & $-0.0094$  & $ 0.2283$   & $ 0.0062$    & $ 0.0251$
\\
            &     &  s.d  & $ 0.0630$  & $ 0.0883$   & $ 0.0851$    & $ 0.1081$
\\
            &     &  MSE  & $ 0.0041$  & $ 0.0599$   & $ 0.0073$    & $ 0.0123$
\\ \hline
            & 100 &  mean & $-0.0048$  & $-0.4166$   & $-0.0449$    & $-0.0871$
\\
            &     &  bias & $ 0.0048$  & $ 0.4166$   & $ 0.0449$    & $ 0.0871$
\\
            &     &  s.d  & $ 0.1763$  & $ 0.2215$   & $ 0.1620$    & $ 0.1811$
\\
            &     &  MSE  & $ 0.0311$  & $ 0.2226$   & $ 0.0283$    & $ 0.0404$
\\
\cline{2-7}
            & 300 &  mean & $-0.0122$  & $-0.3230$   & $-0.0273$    & $-0.0426$
\\
$d_X=1.0, d_W=0.0$       &     &  bias & $ 0.0122$  & $ 0.3230$   & $ 0.0273$
& $
0.0426$ \\
            &     &  s.d  & $ 0.1076$  & $ 0.1296$   & $ 0.1094$    & $ 0.1277$
\\
            &     &  MSE  & $ 0.0117$  & $ 0.1211$   & $ 0.0127$    & $ 0.0181$
\\
\cline{2-7}
            & 800 &  mean & $ 0.0059$  & $-0.2181$   & $-0.0107$    & $-0.0222$
\\
            &     &  bias & $-0.0059$  & $ 0.2181$   & $ 0.0107$    & $ 0.0222$
\\
            &     &  s.d  & $ 0.0648$  & $ 0.0823$   & $ 0.0629$    & $ 0.0909$
\\
            &     &  MSE  & $ 0.0042$  & $ 0.0544$   & $ 0.0041$    & $ 0.0088$
\\
            \hline\hline
\end{tabular}
\label{tab:diff}
\end{center}}
\end{table}

\section{Application}\label{S:application}
\begin{figure}[!ht]
\includegraphics[viewport=0 0 450 250,scale=0.8,clip]{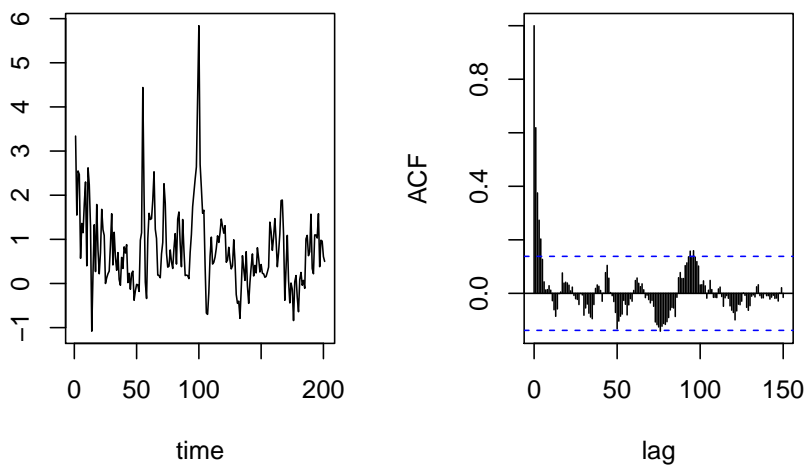}
\caption{IGP-DI series and its sample autocorrelation function: period from Aug/94 to
Apr/11.}
\label{fig:igpdi}
\end{figure}

IGP-DI is the general price index with domestic availability and is calculated by 
Funda\c c\~ao Get\'ulio Vargas, Brazil. The series comprises monthly observations
from August 1994 to April 2011 (total of 201 observations). The series and its ACF are 
displayed in  Figure \ref{fig:igpdi}. The observations of the  months February  1999 
($4.44\%$), October 2002 ($4.21\%$) and November 2002 ($5.84\%$) are  possibly  
outliers. Looking at the plots in Figure \ref{fig:igpdi}, these  suggest that the 
series is stationary and possess long-memory behavior. From the data and using the 
methodologies previously discussed, the parameter $d$ is estimated and the estimates are
displayed in Table \ref{tab:igpdi}. For this application, the estimates $d$ are computed
from the original data (OD)  and from the  modified data (MD)  where the observations of
February 1999, October 2002  and November 2002  are replaced by the sample mean of the 
series. This analysis is a simple exercise to verify the robustness of the estimators in
a real application and, also, to investigate whether the data contains outliers. The 
$d'$ estimates of   OD and MD series are given, respectively, on the left and the right
side of Table \ref{tab:igpdi}. These estimates were calculated using different 
bandwidths in Equation \ref{eqeReg}($m' = n^\alpha$) and $\beta$ was  fixed as in the 
simulation study. In both series, for a fixed $\alpha$, the robust methods present 
similar results. The estimates maintain the same empirical property across the bandwidth
values. In contrast to the robust methods, the classical GPH estimator gives estimates 
that dramatically change from OD to MD data  showing that the observations replaced by 
the mean are possible atypical data.

\begin{table}[tbh]
\begin{center}\caption{{Estimates of $d$: IGP-DI data, period from Aug/94 to Apr/11.}}
{\footnotesize
\begin{tabular}{l|c|c|c|c|c|c|c|c}
\hline\hline
& \multicolumn{4}{c}{Original time series} & \multicolumn{4}{|c}{Modified time series}
\\ \hline
Estimator & $\alpha=0.5$ & $\alpha=0.6$ & $\alpha=0.7$ & $\alpha=0.8$ & $%
\alpha=0.5$& $\alpha=0.6$ & $\alpha=0.7$ & $\alpha=0.8$ \\ \hline
$d_{GPH}$      & 0.0757 & 0.1205 & 0.3431 & 0.3759 & 0.3110 & 0.3116 & 0.3713 & 0.3875 \\
               &(0.3417)&(0.1869)&(0.1389)&(0.0888)&(0.1586)&(0.1077)&(0.0909)&(0.0683) \\
$d_{GPHR_P}$   & 0.1802 & 0.2335 & 0.2269 & 0.2397 & 0.1630 & 0.2077 & 0.2078 & 0.2230 \\
	       &(0.0857)&(0.0745)&(0.0469)&(0.0331)&(0.0782)&(0.0603)&(0.0385)&(0.0251) \\
$d_{GPHR_{TH}}$& 0.1718 & 0.1919 & 0.2125 & 0.2379 & 0.1545 & 0.1782 & 0.1968 & 0.2231 \\
	       &(0.0742)&(0.0508)&(0.0303)&(0.0210)&(0.0673)&(0.0436)&(0.0259)&(0.0170) \\
$d_{GPHR_B}$   & 0.1522 & 0.1788 & 0.2047 & 0.2327 & 0.1379 & 0.1667 & 0.1896 & 0.2181 \\
	       &(0.0641)&(0.0433)&(0.0262)&(0.0183)&(0.0586)&(0.0378)&(0.0227)&(0.0151) \\
$d_{GPHR}$     & 0.1662 & 0.2628 & 0.2454 & 0.2285 & 0.1500 & 0.2211 & 0.2215 & 0.2228 \\
	       &(0.0862)&(0.0995)&(0.0671)&(0.0436)&(0.0794)&(0.0717)&(0.0511)&(0.0328) \\ \hline\hline
\end{tabular}%
}
\end{center}
\label{tab:igpdi}
\end{table}


\section{Concluding remarks and future direction}
This paper investigates the effect of outliers in the estimation of the fractional 
parameter $d$ in   the ARFIMA($p,d,q$) model and, also, discusses the asymptotical and 
empirical properties of the robust autocovariance and spectral estimators, previously 
given in \citeasnoun{fajardo:reisen:cribari:2009} and 
\citeasnoun{levy:boistard:moulines:taqqu:reisen:2011a},  for the case of time series 
with short and long-memory properties.  These studies support the use of the robust 
estimators to estimate the long-memory parameter when Gaussian long-memory time series 
are contaminated with additive outliers. Non-stationary time series with outliers are 
also studied and the investigation reveals that the robust method can be used as an 
alternative estimation procedure in  time series with fractional differences.
As previously stated, the asymptotical properties of the robust  estimator under the 
study  still remain to be investigated.  The robust ACF method discussed here has also 
been  used in other contexts such as in the estimation of periodic process 
(\citeasnoun{sarnaglia:reisen:levy:2010}) and in seasonal ARFIMA processes (this is 
one of the current research of the authors).

%
%
%

\bibliographystyle{agsm}
\bibliography{/media/FAJARDO/Pesquisa/Articulos/references/references}

%
%
%
%
%

\end{document}